\documentclass[prl,twocolumn,preprintnumbers,amsmath,amssymb,superscriptaddress]{revtex4}

\usepackage{graphicx}% Include figure files
\usepackage{dcolumn}% Align table columns on decimal point
\usepackage{bm}% bold math
\usepackage{mathrsfs}
\usepackage{subfigure}
\usepackage{natbib}
\usepackage{amsmath}
\usepackage{amssymb}
\usepackage[english]{babel}

\begin{document}
\title{Band Hybridization Induced Odd-Frequency Pairing in Multiband Superconductors}

\author{L. Komendov\'{a}}
\affiliation{Department of Physics and Astronomy, Uppsala University, Box 530, SE-751 21 Uppsala}

\author{A. V. Balatsky}
\affiliation{NORDITA, Center for Quantum Materials, KTH Royal Institute of Technology, and Stockholm University, Roslagstullsbacken 23, SE-106 91 Stockholm, Sweden}
\affiliation{Institute for Materials Science, Los Alamos National Laboratory, Los Alamos, NM 87545, United States}

\author{A. M. Black-Schaffer}
\affiliation{Department of Physics and Astronomy, Uppsala University, Box 530, SE-751 21 Uppsala}


\date{\today}

\begin{abstract}
We investigate how hybridization (single-quasiparticle scattering) between two superconducting bands induces odd-frequency superconductivity in a multiband superconductor. An explicit derivation of the odd-frequency pairing correlation and its full frequency dependence is given. We find that the density of states is modified, from the sum of the two BCS spectra, at higher energies by additional hybridization gaps with strong coherence peaks when odd-frequency pairing is present.
\end{abstract}

\maketitle
% --------------------------------------- %
% INTRODUCTION:
% --------------------------------------- %
% Multiband materials:
Many materials have multiple bands close to the Fermi level.
It is then not surprising that there also exist many superconductors with more than one superconducting band.
Well-known multiband superconductors are MgB$_2$ \cite{Nagamatsu01,Choi02, Souma03}, which hosts two distinct superconducting gaps, and the iron-based superconductors \cite{Kamihara08, Hunte08, Stewart11}, where the order parameter changes sign between different bands \cite{Mazin08, Hanaguri10}. Multiband superconductivity has also been suggested to be important in simple metals \cite{Shen&Phillips65, Boaknin03}, heavy fermion compounds \cite{Stewart01, Jourdan04, Rourke05, Seyfarth05, Hill08}, different carbides \cite{Bergk08, Kuroiwa08}, and Chevrel phases \cite{Petrovic11}, as well as for engineering time-reversal invariant topological superconducting states \cite{Deng12, Deng13}.

% Josephson coupling:
Multiple superconducting bands allow for unusual coupling effects.
Early, and much current, theoretical focus has been oriented towards studying the effects of Josephson coupling between different superconducting bands, i.e.~the exchange of whole Cooper pairs between bands \cite{Suhl, Zhitomirsky, Chaves, Komendova, Babaev, Carlstrom}.
% Single-particle scattering:
However, conceptually much simpler is single-quasiparticle scattering, or tunneling, between the superconducting bands.
The origin of interband quasiparticle scattering can be impurity scattering, but a common intrinsic source is superconductivity associated with specific orbitals, which subsequently hybridize to form the low-energy bands around the Fermi surface.
Tunneling spectroscopy has found significant effects of interband quasiparticle scattering in silicon clathrate Ba$_8$Si$_{46}$ \cite{Noat1, Noat2}, iron-based superconductors \cite{Noat-Fe}, and MgB$_2$ \cite{Schmidt}. Recent ARPES data on MgB$_2$ have also shown that interband scattering due to disorder can be important \cite{Mou}. 

% Effects of single-particle scattering:
% Hybridization:
Theoretically, interband single-quasiparticle scattering results in both band hybridization and {\it interband} pairing, where the two electrons forming a Cooper pair belong to {\it different} bands. Straightforward effects of band hybridization were studied already quite early on \cite{Schopohl, McMillan, Japiassu92}. More recently, band hybridization has also been proposed to influence the nodal structure of the superconducting gap in iron-based superconductors \cite{Moreo09, Hinojosa, Cvetkovic, Ong&Coleman}.
% Interband pairing:
On the other hand, consequences of the induced interband pairing have been much less discussed.
Pure interband pairing in cold atoms and quantum chromodynamics systems has been proposed to give a ``breached" regime containing both a superfluid and a normal liquid \cite{Liu&Wilczek03, Gubankova&Wilczek03}, however, in band-hybridized superconductors the interband pairing is also accompanied by (usually large) conventional intraband pairings.
Still, it was recently pointed out, using symmetry arguments and simple mean-field BCS calculations, that odd-frequency, odd-interband pairing can be ubiquitous in multiband superconductors \cite{AnnicaMB}.

%Odd-w:
The fermionic nature of the superconducting wave function usually renders a division into spin-singlet even-parity (i.e.~$s$-, $d$-wave) or spin-triplet odd-parity ($p$-wave) superconductors, but it can also be even or odd under time, or equivalently frequency \cite{Berezinskii,Balatsky92,odd-bulk}. Examples of this are odd-frequency spin-triplet $s$-wave superconductivity giving rise to long-scale proximity effects in superconductor-ferromagnet systems \cite{Bergeret, reviewBergeret, Aperis} or odd-frequency spin-singlet $p$-wave pairing in non-magnetic junctions \cite{Tanaka2007b, Tanaka2007,Tanaka2007PRB}. In multiband superconductors the band index offers the additional possibility of spin-singlet $s$-wave pairing which is odd in both band and frequency, without translation or spin rotation symmetry breaking.

% What we do:
In this work we derive the exact Green's functions for a generic multiband superconductor with single-quasiparticle scattering. We thus obtain the full frequency dependence of the interband pairing and find that the odd-interband pairing has odd frequency dependence, while the even-interband pairing is even in frequency. By studying the density of states (DOS) we also discover that finite band hybridization gives rise to an extra gap located beyond the original gap edges. This hybridization-induced gap is not fully depleted, but has very pronounced BCS-like coherence peaks. Moreover, the hybridization gap disappears whenever the odd-frequency interband pairing is zero and is thus a directly measurable signal of odd-frequency superconductivity.

% ------------------ %
% RESULTS:
% ------------------ %
To model a generic multiband superconductor we consider two bands, band 1 and 2, with dispersions $\epsilon_{k1}$ and $\epsilon_{k2}$, where $\epsilon_k = \epsilon_{-k}$. For simplicity, we assume that the two bands independently develop conventional spin-singlet $s$-wave superconducting order parameters $\Delta_1$ and $\Delta_2$, respectively, but one of them can also be zero. Finally, we add a small single-quasiparticle hybridization, or scattering, term proportional to $\Gamma$ between the two bands, resulting in the Hamiltonian:
\begin{align}
\label{hamiltonian}
H & = \sum_{k\sigma} \epsilon_{k1} a_{k\sigma}^\dagger a_{k\sigma} + \epsilon_{k2} b_{k\sigma}^\dagger b_{k\sigma} + \sum_{k\sigma} \Gamma(k) a_{k\sigma}^\dagger b_{k\sigma} + {\rm H.c.} \nonumber \\
& + \sum_{k} \Delta_1(k)a^\dagger_{k \uparrow} a^\dagger_{-k \downarrow} + \Delta_2(k) b^\dagger_{k \uparrow} b^\dagger_{-k \downarrow} + {\rm H.c.},
\end{align}
where $a$ $(b)$ is the annihilation operator in band 1 (2).
Alternatively, the band hybridization can be interpreted as a coupling process in real space, with the $a$- and $b$-electrons living to the left and the right of a junction or in different layers \cite{McMillan, Parhizgar14}. In this picture the Josephson coupling would instead be a two-particle tunneling term (not included here).

\begin{figure}
\includegraphics[width=0.8\linewidth]{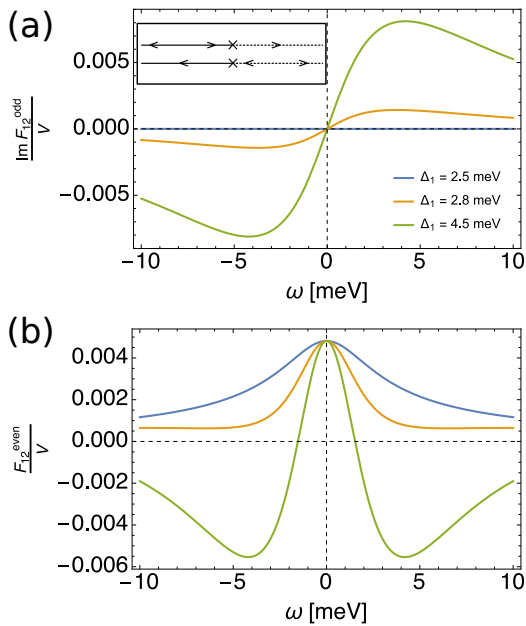}
\caption{Odd- (a) and even- (b) interband pairing amplitudes in meV per nm$^3$ when $\Delta_2 = 2.5$ meV, $\Delta_1 =$ 2.5, 2.8, 4.5 meV (blue, orange, green), and $\Gamma = 3$ meV, with the band structure specified in the main text.
Inset: first order hybridization contributions to the interband pairing; $F_1 \Gamma G_2$ (top), $-\protect \overleftarrow{G}_1 \Gamma F_2$ (bottom). Solid (dashed) line represents the propagator in band 1 (2) and $\times$ the hybridization.
\label{fig1}}
\end{figure}

We start by calculating the spin-singlet $s$-wave interband anomalous Green's functions $F_{12}$ and $F_{21}$, which express the pairing correlations of two electrons belonging to different bands. Assuming $\Gamma$ to be a small parameter, we can use standard perturbation theory \cite{Mahan-book}. The first order contributions are then represented by the schematic diagrams in the inset of Fig.~\ref{fig1}(a) and give:
$F_{12}^{(1)} = F_1 \Gamma G_2 - \overleftarrow{G}_1 \Gamma F_2$,
where $\overleftarrow{G}$ is the hole propagator, i.e., $\protect \overleftarrow{G} = -G(-k, -\omega).$ The minus sign before the second term is due to scattering of hole propagators (left-going arrows). 
The normal and anomalous propagators without the hybridization are as usual~\cite{Mahan-book}:
\begin{equation}
 \label{matrixGF}
 \begin{pmatrix}
    G_j & F_j \\  F_j^\dagger & \overleftarrow{G}_j     \\      \end{pmatrix} =
    \frac{1}{(i\omega)^2 - E_{j}^2}
  \begin{pmatrix}
    i \omega + \epsilon_{kj} & \Delta_j \\        \Delta_j^* &  i \omega - \epsilon_{kj} \\      \end{pmatrix},
 \end{equation}
where $E_j^2 = E_{kj}^2 = \epsilon_{kj}^2 + |\Delta_j|^2$ and $\omega = \omega_n = \pi (2n+1) k_B T$ are the fermionic Matsubara frequencies.
Using these expressions we arrive at
$F_{12}^{(1)} = \Gamma [i\omega(\Delta_1 - \Delta_2) + \Delta_1 \epsilon_{k2} + \Delta_2 \epsilon_{k1}]/[(\omega^2 + E_{1}^2)(\omega^2 + E_{2}^2)]$.
The next to leading order terms for $F_{12}$ are cubic in $\Gamma$. Further organizing the perturbation expansion of the interband pairing of a given order $n$ in a systematic way, we find several recursion relationships \footnote{See Supplementary Material for a derivation.}. These can be compactly written in a matrix form as:
 \begin{equation}
 \label{matrixGS}
 \begin{pmatrix}
    \overleftarrow{G}_{12}^{(n)} \\ F_{12}^{(n)}         \\      \end{pmatrix} = g
  \begin{pmatrix}
    e & f \\        -f^* & e^* \\      \end{pmatrix}
 \begin{pmatrix}
   \overleftarrow{G}_{12}^{(n-2)} \\ F_{12}^{(n-2)}        \\      \end{pmatrix},
 \end{equation}
 where $g = \Gamma^2 / [ (\omega^2 + E_{k1}^2)(\omega^2 + E_{k2}^2)  ] $, $e = (i \omega - \epsilon_1)(i \omega - \epsilon_2) -\Delta_1 \Delta_2^*$ and $f = -i \omega(\Delta_1^* - \Delta_2^*) + \Delta_1^*\epsilon_2 + \Delta_2^*\epsilon_1$. For the starting point of the recursion we use the first order anomalous interband Green's function $F_{12}^{(1)}$ and the corresponding normal propagator:
$\overleftarrow{G}_{12}^{(1)} = \Gamma [\Delta_1\Delta_2^* -(i \omega - \epsilon_1)(i \omega - \epsilon_2)]/[(\omega^2 + E_{1}^2)(\omega^2 + E_{2}^2)]$.
From Eq.~\eqref{matrixGS} we recognize that the Green's functions to infinite order can be written as a geometric series, where the quotient is a two-by-two matrix. The formal criterion for a matrix geometric series to be convergent is that the norm (i.e.~the largest singular value) of the coefficient matrix is $<1$. From this we arrive at the condition
$\Gamma \lesssim (\omega^2 + E_1^2)^{1/4} (\omega^2 + E_2^2)^{1/4}$,
which translates into:
\begin{equation}
\label{convergence}
\Gamma \lesssim \sqrt{|\Delta_1||\Delta_2|},
\end{equation}
meaning that the series is always convergent for sufficiently small $\Gamma$, provided that $\Delta_1$ and $\Delta_2$ are both finite. This allows us to sum the infinite series and we arrive at
$F_{12} =  \Gamma [i\omega(\Delta_1-\Delta_2) +\Delta_2\epsilon_1 + \Delta_1 \epsilon_2]/D$,
where
$D = (\omega^2 +E_1^2)(\omega^2 + E_2^2)
-\Gamma^2 [ 2(\epsilon_1 \epsilon_2 -\omega^2) -\Delta_2^*\Delta_1 - \Delta_1^*\Delta_2 ] + \Gamma^4$.
The expression for $F_{21}$ is obtained by exchanging the band indices and we can also form the odd and even combinations of $F_{12}$ and $F_{21}$ with respect to the band index:
\begin{align}
\label{delta-odd}
F_{12}^{\mathrm{odd}}(\mathbf{k},i\omega) & = \frac{F_{12} - F_{21}}{2} = i \omega \Gamma (\Delta_1 - \Delta_2)/D \\
\label{delta-even}
F_{12}^{\mathrm{even}}(\mathbf{k},i\omega) & = \frac{F_{12} + F_{21}}{2} = \Gamma (\Delta_1 \epsilon_{k2} + \Delta_2 \epsilon_{k1})/D.
\end{align}
The odd-band combination is directly seen to be odd in frequency, whereas the even-band combination has a conventional even-frequency dependence. This is fully consistent with Fermi-Dirac statistics for spin-singlet $s$-wave superconducting pairing.
Furthermore, we see that interband pairing always requires a finite band hybridization and that the odd-interband pairing also requires $\Delta_1 \neq \Delta_2$.
Equations~\eqref{delta-odd}-\eqref{delta-even} can be Fourier transformed to real space and then evaluated numerically, with the result shown in Fig.~\ref{fig1}. By definition, the odd-frequency pairing amplitude must be zero at $\omega = 0$. Close to $\omega = 0$, we find that the leading term in $F^{\mathrm{odd}}$ is linear in $\omega$. However, the slope can initially be large and then abruptly change sign to asymptotically go to zero for large $\omega$, resulting in an approximate $1/\omega$-dependence, which has also been found for some odd-frequency states \cite{AnnicaTI, Coleman}. 
%
% FIGURE 1:

By forming an analogue of Eq.~\eqref{matrixGS} for the intraband anomalous Green's function in band 1 we find
$F_1 = \{ \Delta_1 [(i \omega)^2 -E_2^2] -\Gamma^2 \Delta_2 \}/D$
and the corresponding normal Green's function:
$G_1 = \{ (i\omega + \epsilon_1) [(i \omega)^2 -E_2^2] -\Gamma^2 (i\omega - \epsilon_2) \}  /D$.
The expressions for $F_2$ and $G_2$ are obtained by mutually exchanging band indices. The expressions for the normal Green's functions allow us to compute the DOS using the standard formula:
\begin{equation}
N(E) = -\frac{1}{\pi} \mathrm{Im} \, \mathrm{Tr} \, G(E+i\delta), \ \ \delta  \rightarrow 0^{+}.
\end{equation}
Here the trace involves a sum over band and spin indices and an integral over $k$-space. We use similar expressions, but unsummed over the band index, to define the partial DOS $N_1$ and $N_2$.

\begin{figure}
\includegraphics[width=\linewidth]{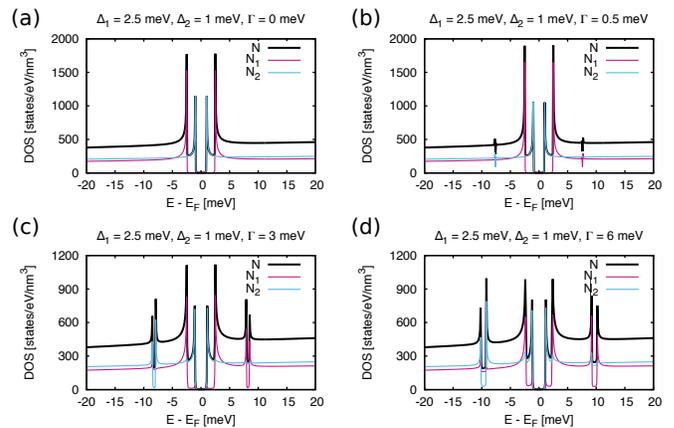}
\caption{Total ($N$) and partial densities of states ($N_1$, $N_2$) when $\Delta_1 = 2.5$ meV and $\Delta_2 = 1$ meV for different values of  $\Gamma = 0, 0.5, 3, 6$~meV (a-d), with the band structure specified in the main text.}
\label{fig2}
\end{figure}

\begin{figure}
\includegraphics[width=\linewidth]{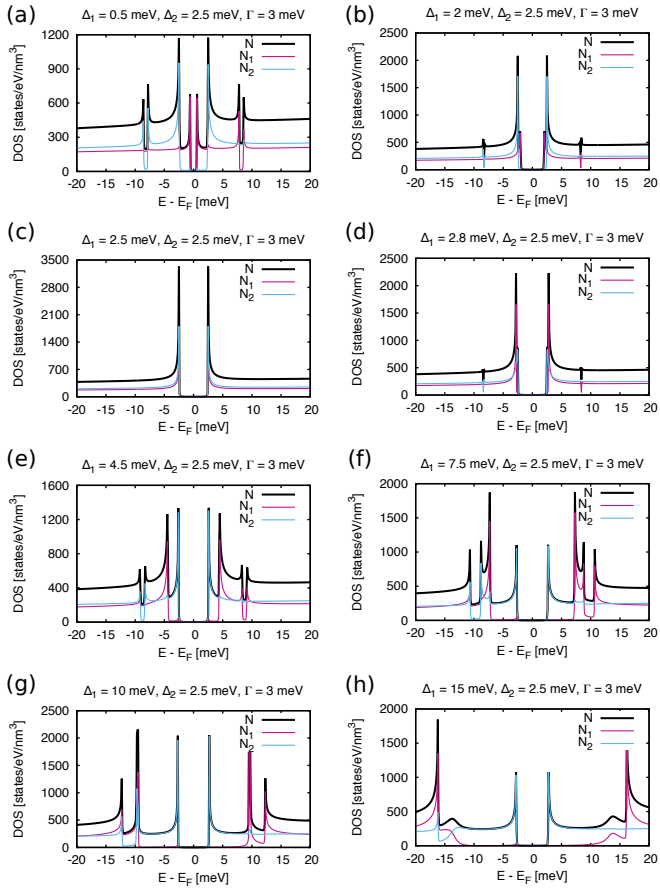}
\caption{Total ($N$) and partial densities of states ($N_1$, $N_2$) when $\Delta_2 = 2.5$ meV and $\Gamma = 3$ meV for different values of $\Delta_1 =  0.5, 2, 2.5, 2.8, 4.5, 7.5, 10, 15$~meV (a-h), with the band structure specified in the main text.}
\label{fig3}
\end{figure}

\begin{figure}
\includegraphics[width=\linewidth]{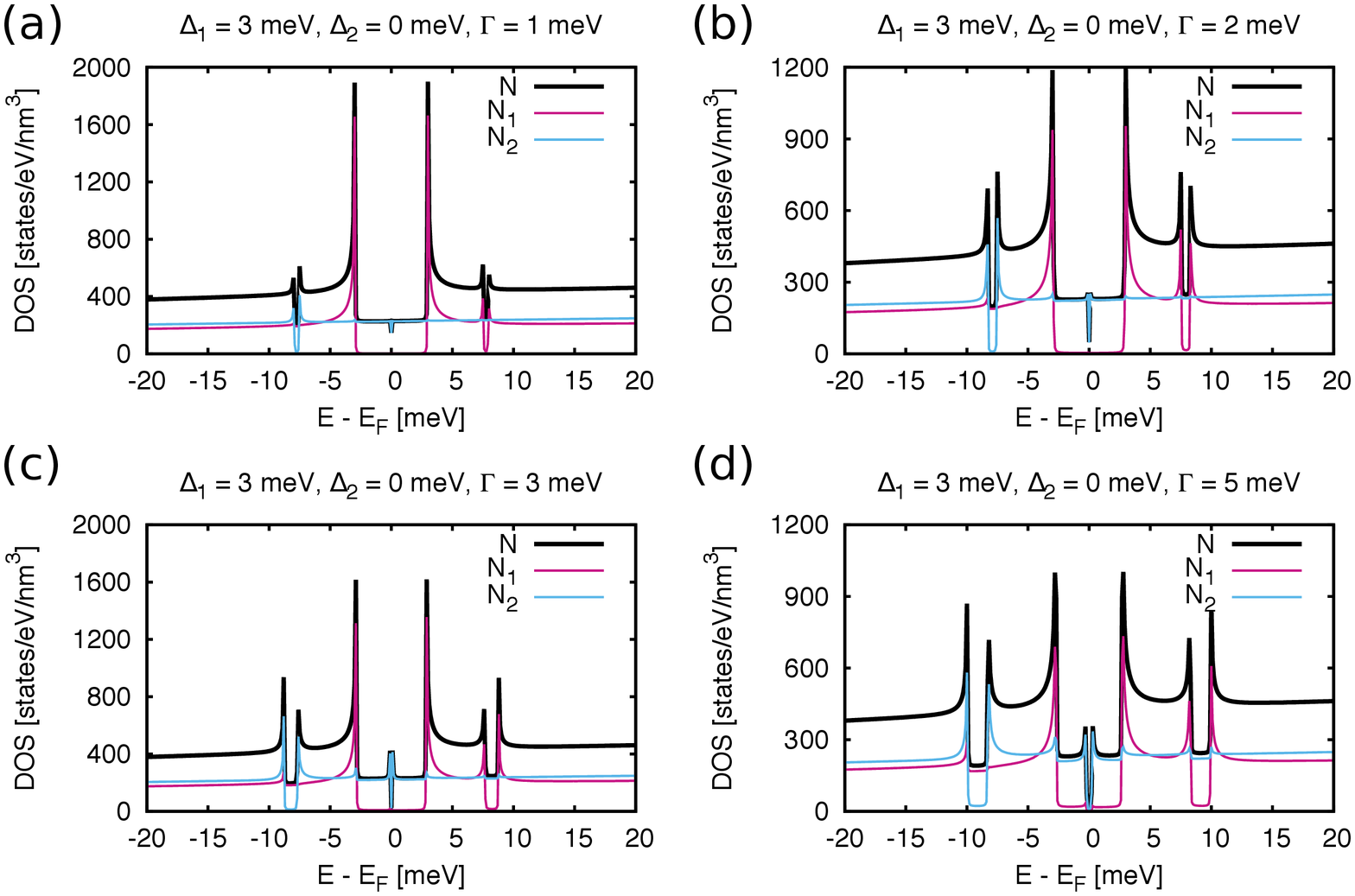}
\caption{Total ($N$) and partial densities of states ($N_1$, $N_2$) when $\Delta_1 = 3$ meV and $\Delta_2 = 0$, i.e.~only one band with native superconductivity, for different values of $\Gamma =  1, 2, 3, 5$~meV (a-d) with the band structure is specified in the main text.}
\label{fig4}
\end{figure}

The total and partial DOS offer a direct connection to experimental measurements on multiband superconductors.
In Figs.~\ref{fig2} - \ref{fig4} we show numerically obtained results for the DOS, explicitly exploring the effect of interband pairing. To most clearly illustrate the effect of interband pairing we use two generic parabolic bands: $\epsilon_j = \hbar^2k^2/2m_j - \mu_j$, with effective masses $m_1= 20 m_e$, $m_2 = 22 m_e$ and distances from the bottom of the bands to the Fermi level $\mu_1 = $ 100~meV and $\mu_2 = $ 105~meV. For this and other band structures we studied, the interband effects are most clearly visible when $\Gamma$ is comparable to $|\epsilon_1 - \epsilon_2|$ for fixed $k \approx k_F$.
The contributions to the DOS are obtained by numerical integration in the range including both Fermi surfaces and we use a smearing parameter $\delta = 0.01$ meV.  We have also independently verified the DOS results by exact diagonalization of the Hamiltonian in Eq.~\eqref{hamiltonian}. In fact, we find a perfect agreement even far beyond the theoretical condition for convergence in Eq.~\eqref{convergence}.

We here especially showcase that unusual features in the DOS are only seen when the conditions $\Gamma \neq 0$ and $\Delta_1 \neq \Delta_2$ are both satisfied, which are exactly the two key criteria for odd-frequency pairing, see Eq.~\eqref{delta-odd}. First, the DOS without any band hybridization, as shown in Fig.~\ref{fig2}(a), is just a sum of two BCS spectra with energy gaps $E_{g1} =\Delta_1$ and $E_{g2} = \Delta_2$, respectively, as expected.
%
% FIGURE 2:
However, when we turn on hybridization, see Figs.~\ref{fig2}(b-d), we see extra, very notable, dips in the DOS located beyond the original gap edges $E_{g1,2}$. These dips, symmetric around zero energy, clearly resemble superconducting gaps with their pronounced BCS-like coherence peaks. However, the DOS in the gap regions are not zero, but instead equal to the partial DOS at these energies. Still, we refer to these features as hybridization-induced gaps.
The hybridization-induced gaps grow in size and move to higher energies for larger band hybridizations, and we associate these features with interband superconducting pairing.
We note that beyond these hybridization-induced gaps, we see no other distinctive features associated with interband pairing. Notably, there are no zero-energy or subgap states, otherwise often associated with odd-frequency pairing \cite{Balatsky92, Tanaka2007, Tanaka2007PRB, Yokoyama07, Dahal09, Asano13, reviewTanaka}. Recent works have pointed out that zero-energy states do not always accompany odd-frequency pairing \cite{AnnicaTISC, AnnicaTI, AnnicaMB, Linder&Robinson}, and odd-frequency, odd-interband pairing provides another example when this is not the case.

In Fig.~\ref{fig3} we instead fix $\Gamma = 3$~meV, and $\Delta_2 = 2.5$~meV but vary $\Delta_1$.
%
% FIGURE 3:
Distinct hybridization-induced gaps are present at energies beyond the original gap edges, independent on the relative size of the two original gaps. The only exception is exactly when $\Delta_1 = \Delta_2$, then the hybridization-induced gaps completely disappear, despite the finite band hybridization, see Fig.~\ref{fig3}(c). Detuning the value of $\Delta_1$ slightly from that of $\Delta_2$ results in small, but noticeable, dips in the DOS, at the positions where the full gaps develop for increasing differences between $\Delta_1$and $\Delta_2$.
We thus find that the hybridization-induced gaps are only present in the DOS when both $\Gamma \neq 0$ and $\Delta_1 \neq \Delta_2$. These are exactly the two key criteria for odd-frequency pairing, as seen in Eq.~\eqref{delta-odd}.
In fact, only the odd-frequency interband pairing disappears at $\Delta_1 = \Delta_2$, the even-frequency part is in general non-zero as soon as $\Gamma \neq 0$. Specifically, the even- and odd-frequency interband pairing amplitudes corresponding to the parameters in Fig.~\ref{fig3}(c)-(e) are plotted in Fig.~\ref{fig1}. From there it is clear that the even-frequency interband pairing is large and not changing significantly around $\omega = 0$, while the odd-frequency part changes from identically zero in Fig.~\ref{fig3}(c) to a notable non-zero derivative at $\omega = 0$ for Figs.~\ref{fig3}(d)-(e).
We can thus conclude that odd-frequency interband pairing is necessary for producing the hybridization-induced gaps. Detecting gaps beyond the original two superconducting gaps in multiband superconductors is therefore a clear sign of the presence of odd-frequency pairing.
Intriguingly, additional gap features have already been reported in the multiband superconductor Ba$_8$Si$_{46}$ \cite{Noat1}.

Finally, we also show that only one band has to be natively superconducting for the hybridization-induced gaps to be present. In Fig.~\ref{fig4} we display how the hybridization-induced gap grows with increasing $\Gamma$ when $\Delta_2 = 0$.
%
% FIGURE 4:
Finite single-particle hybridization $\Gamma$ results in a proximity-induced gap also in the second band, which is manifested as a gap around zero energy, although always smaller than $E_{g1}  = \Delta_1$.
For all parameter choices in Fig.~\ref{fig4} the spectrum always has a gap at zero energy, i.e.~there are no zero-energy states. This is not clearly seen in Fig.~\ref{fig4}(a) due to the finite smearing parameter, but it is clearly visible in exact diagonalization results which do not suffer from the same problem.
In addition, a finite $\Gamma$ results in finite odd-frequency interband pairing, and we consequently also see hybridization-induced gaps beyond the $\Delta_1$ gap, which also grow with $\Gamma$. In fact, these gaps are much more pronounced than the proximity-induced gap in band 2 at zero energy.

% SUMMARY:
In summary, we have studied the effect of single-quasiparticle hybridization or scattering in a two-band superconductor.
By performing perturbation theory to infinite order in the hybridization term, we have obtained the exact, fully frequency dependent, expression for the interband pairing, which can be divided up into odd-frequency odd-interband and even-frequency even-interband pairing.
The conditions for finite odd-frequency interband pairing are (a) finite single-quasiparticle hybridization and (b) a non-zero difference between the original superconducting gaps; no applied magnetic field, inhomogeneity, or interface is required.
Furthermore, we have shown that the DOS develops non-trivial gaps features with distinct coherence peaks beyond the original gap edges only if the conditions for odd-frequency pairing are satisfied, otherwise the spectrum just a sum of two BCS spectra. Detecting such additional gaps thus provides experimental evidence of odd-frequency pairing in multiband superconductors.

\begin{acknowledgments} 
We would like to thank to K.~Bj\"ornson, D.~Kuzmanovski, T.~L\"othman, and S.~Nakosai for valuable discussions.
We acknowledge funding from the Wenner-Gren Foundations, the Swedish Research Council (Vetenskapsr\aa det), the G\"oran Gustafsson Foundation, and the Swedish Foundation for Strategic Research (SSF) (LK and ABS), and the European Research Council (ERC) DM-321031 (AVB). Work at Los Alamos was supported by the US DOE Basic Sciences E 304 for the National Nuclear Security Administration of the US Department of Energy under Contract No.~DE-AC52-06NA25396.
\end{acknowledgments}

% BIBLIOGRAPHY:

%%%%%%%%%% Merge with supplemental materials %%%%%%%%%%
\pagebreak
\widetext
\begin{center}
\textbf{\large Band Hybridization Induced Odd-Frequency Pairing in Multiband Superconductors: \\ Supplementary material}
\end{center}
\twocolumngrid
%%%%%%%%%% Merge with supplemental materials %%%%%%%%%%
%%%%%%%%%% Prefix a "S" to all equations, figures, tables and reset the counter %%%%%%%%%%
\setcounter{equation}{0}
\setcounter{figure}{0}
\setcounter{table}{0}
\setcounter{page}{1}
\makeatletter
\renewcommand{\theequation}{S\arabic{equation}}
\renewcommand{\thefigure}{S\arabic{figure}}
\renewcommand{\bibnumfmt}[1]{[S#1]}
\renewcommand{\citenumfont}[1]{S#1}
%%%%%%%%%% Prefix a "S" to all equations, figures, tables and reset the counter %%%%%%%%%%

In this Supplementary Material we derive in detail the exact result (within perturbation theory) of the interband pairing amplitude in a two-band superconductor with hybridized bands.

The Hamiltonian for a generic multiband superconductor can be written as (same as Eq.~(1) in the main text):
\begin{align}
H & = \sum_{k\sigma} \epsilon_{k1} a_{k\sigma}^\dagger a_{k\sigma} + \epsilon_{k2} b_{k\sigma}^\dagger b_{k\sigma} + \sum_{k\sigma} \Gamma(k) a_{k\sigma}^\dagger b_{k\sigma} + {\rm H.c.} \nonumber \\
& + \sum_{k} \Delta_1(k)a^\dagger_{k \uparrow} a^\dagger_{-k \downarrow} + \Delta_2(k) b^\dagger_{k \uparrow} b^\dagger_{-k \downarrow} + {\rm H.c.}.
\end{align}
It describes two superconducting bands coupled by a hybridization, or scattering,  term $\Gamma(k)$, which we treat as a perturbation. We thus start with two copies of the single band superconducting Green's functions (same as Eq.~(2) in the main text): 
\begin{equation}
 \begin{pmatrix}
    G_j & F_j \\  F_j^\dagger & \overleftarrow{G}_j     \\      \end{pmatrix} =
    \frac{1}{(i\omega)^2 - E_{j}^2}
  \begin{pmatrix}
    i \omega + \epsilon_{kj} & \Delta_j \\        \Delta_j^* &  i \omega - \epsilon_{kj} \\      \end{pmatrix},
 \end{equation}
which we use together with the hybridization term to build up the full Green's functions of the system. 
Here $G$ is the normal electron propagator, which we schematically denote by \includegraphics[scale=0.35]{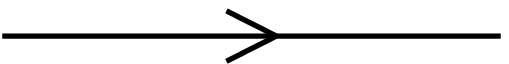}, and $\overleftarrow{G}$ is the hole propagator denoted by  \includegraphics[scale=0.35]{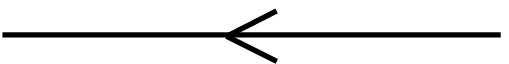}. Moreover, $F$ is the anomalous propagator, denoted by \includegraphics[scale=0.35]{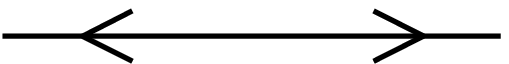}, and $F^\dagger$ is its Hermitian conjugate \includegraphics[scale=0.35]{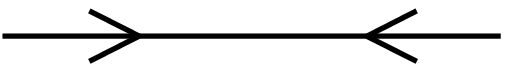}. We will furthermore use solid lines to denote the propagators in band 1 and dashed lines for band 2. The hybridization term is represented by $\times$, which always connects two propagators from different bands with the same direction of the arrow (due to momentum conservation).

Now, we want to calculate the interband pairing amplitude $F_{12}$, which corresponds to the sum of all processes of the type \includegraphics[scale=0.35]{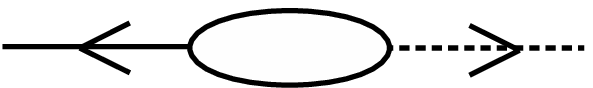}.   
The simplest processes, here indicated by the superscript (1), include just one scattering event $\times$: $F_{12}^{(1)} = \includegraphics[scale=0.35]{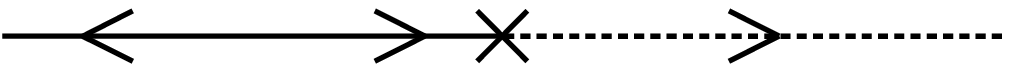} - \includegraphics[scale=0.35]{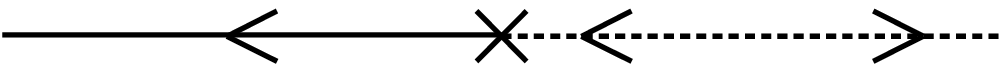} = F_1 \Gamma G_2 - \overleftarrow{G}_1 \Gamma F_2$. 
The sign of a particular process is simply given by $(-1)^l$, where $l$ is the number of hole scattering events, i.e.~scattering connecting two left-going arrows.    

Since $F_{12}$ of any given order has to end with a right-pointing arrow in band two, there are only two possibilities, either it ends with \includegraphics[scale=0.35]{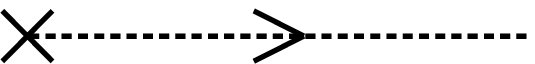} or with \includegraphics[scale=0.35]{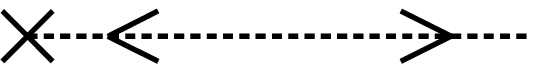}. Similarly, it needs to start with a left-going solid arrow, which also limits the options. By drawing all possible processes and translating them into formulas, we can write: 
\begin{equation}
\label{F12intermediate}
F_{12}^{(n)} = -\overleftarrow{G}_1^{(n-1)} \includegraphics[scale=0.35]{xF} + F_1^{(n-1)} \includegraphics[scale=0.35]{xG}.
\end{equation}
Carrying this procedure one step further, we write $\overleftarrow{G}_1^{(n-1)}$ and $F_1^{(n-1)}$ using $F_{12}^{(n-2)}$ and $\overleftarrow{G}_{12}^{(n-2)}$:
\begin{align}
\overleftarrow{G}_1^{(n-1)} =&	F_{12}^{(n-2)} \includegraphics[scale=0.35]{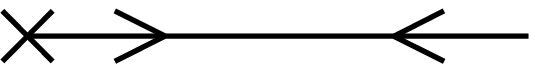} - \overleftarrow{G}_{12}^{(n-2)} \includegraphics[scale=0.35]{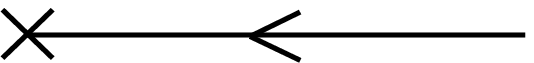}			\\
F_1^{(n-1)} =&  F_{12}^{(n-2)}  \includegraphics[scale=0.35]{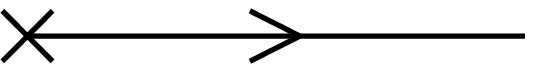} - \overleftarrow{G}_{12}^{(n-2)} \includegraphics[scale=0.35]{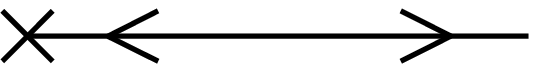}.
\end{align}
 Plugging these into Eq.~\eqref{F12intermediate} we arrive at: 
 \begin{multline}
 F_{12}^{(n)} = \overleftarrow{G}_{12}^{(n-2)} [ \includegraphics[scale=0.25]{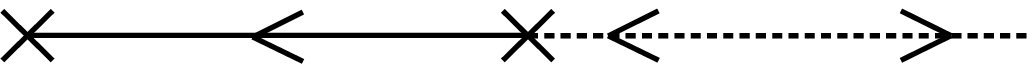}  - \includegraphics[scale=0.25]{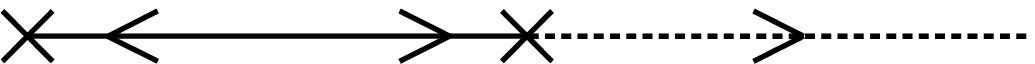}] + \\
 F_{12}^{(n-2)} [\includegraphics[scale=0.25]{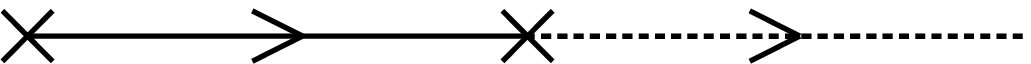} - \includegraphics[scale=0.25]{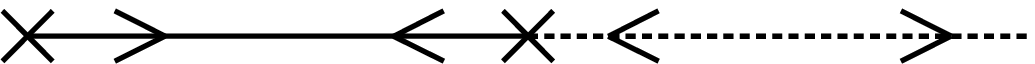} ].
 \end{multline}
 By a similar procedure we get: 
 \begin{multline}
 \overleftarrow{G}_{12}^{(n)} = \overleftarrow{G}_{12}^{(n-2)} [ \includegraphics[scale=0.25]{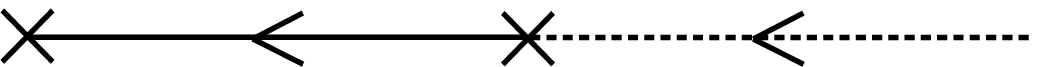} - \includegraphics[scale=0.25]{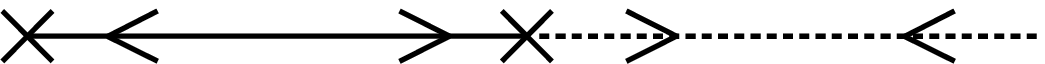}] + \\
 F_{12}^{(n-2)} [\includegraphics[scale=0.25]{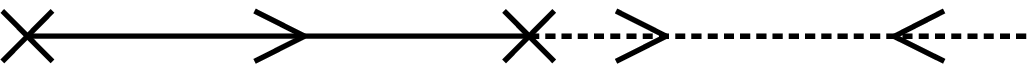} - \includegraphics[scale=0.25]{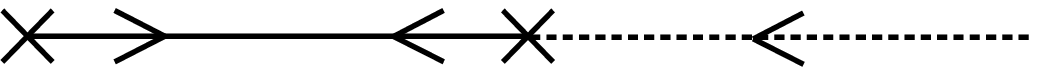} ].
 \end{multline}
 Thus, we get a closed set of equations if we consider $F_{12}$ and $\overleftarrow{G}_{12}$ together, which is easiest done in a matrix formalism. Note that all arrow diagrams can be directly translated to specific formulas using Eq.~\eqref{matrixGF}, e.g. $\includegraphics[scale=0.35]{H1F2}   = - \Gamma \overleftarrow{G}_1 (- \Gamma) F_2 = {(-\Gamma)^2 (i \omega - \epsilon_1)\Delta_2 }/\{[(i \omega)^2 - E_1^2] [(i \omega)^2 - E_2^2]\}$. 

Summarizing, we arrive at the matrix recursion relation in Eq.~(3) in the main text:
\begin{equation}
 \begin{pmatrix}
    \overleftarrow{G}_{12}^{(n)} \\ F_{12}^{(n)}         \\      \end{pmatrix} = g
  \begin{pmatrix}
    e & f \\        -f^* & e^* \\      \end{pmatrix}
 \begin{pmatrix}
   \overleftarrow{G}_{12}^{(n-2)} \\ F_{12}^{(n-2)}        \\      \end{pmatrix},
 \end{equation}
 where $g = \Gamma^2 / [ (\omega^2 + E_{k1}^2)(\omega^2 + E_{k2}^2)  ] $, $e = (i \omega - \epsilon_1)(i \omega - \epsilon_2) -\Delta_1 \Delta_2^*$ and $f = -i \omega(\Delta_1^* - \Delta_2^*) + \Delta_1^*\epsilon_2 + \Delta_2^*\epsilon_1$.

This is a matrix geometric series and is thus easily summed giving:
\begin{equation}
 \begin{pmatrix}
    \overleftarrow{G}_{12} \\ F_{12}         \\      \end{pmatrix} = q  
  \begin{pmatrix}
    1-ge^*  & gf \\        -gf^* & 1-ge \\      \end{pmatrix}
 \begin{pmatrix}
   \overleftarrow{G}_{12}^{(1)} \\ F_{12}^{(1)}        \\      \end{pmatrix},
 \end{equation}
where $q = [(1-ge)(1-ge^*)+g^2 |f|^2]^{-1}$, $\overleftarrow{G}_{12}^{(1)} = \includegraphics[scale=0.35]{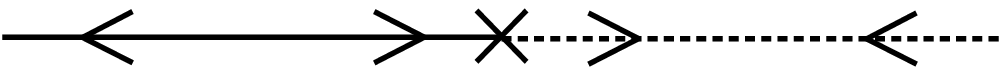} - \includegraphics[scale=0.35]{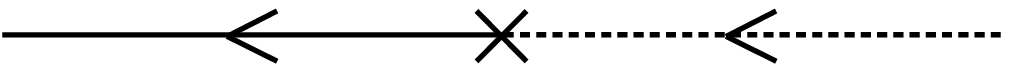}  = \Gamma [\Delta_1\Delta_2^* -(i \omega - \epsilon_1)(i \omega - \epsilon_2)]/[(\omega^2 + E_{1}^2)(\omega^2 + E_{2}^2)]$.
Some terms cancel and we get $F_{12} = qF_{12}^{(1)}$. Finally $F_{12} =  \Gamma [i\omega(\Delta_1-\Delta_2) +\Delta_2\epsilon_1 + \Delta_1 \epsilon_2]/D$,
where
$D = (\omega^2 +E_1^2)(\omega^2 + E_2^2) 
-\Gamma^2 [ 2(\epsilon_1 \epsilon_2 -\omega^2) -\Delta_2^*\Delta_1 - \Delta_1^*\Delta_2 ] + \Gamma^4$.

For the normal propagator $G_1$, needed for calculating the DOS, we create a matrix equation together with $F_1^\dagger$, otherwise the procedure proceeds in the same manner as above.

\end{document}